\documentclass[letterpaper]{article}
\title{Formal methods, Standard Automata Theory, and Process Algebra}
\author{Victor Yodaiken
}

\usepackage{arxiv}
\usepackage{arxiv}
\usepackage[T1]{fontenc}    
\usepackage{graphicx}
\usepackage{caption}
\usepackage{subcaption}
\usepackage{url}
\usepackage{hyperref}
\usepackage{amsfonts}       
\usepackage{amsmath}       
\usepackage{amsthm}       
\newcommand{\ess}{\epsilon}
\newcommand{\xy}{\cdot}
\newcommand{\concat}{\mathop{\mathtt{ concat }}}
\begin{document}


\begin{abstract}
Classical automata theory is far more capable of modeling complex digital
systems than is widely acknowledged in the ``formal methods'' literature.
This paper takes a second look at automata theory methods that were mostly
developed in the 1950s and 1960s to show how they can be applied to problems
of current era specification and verification of systems, including
concurrent systems. The explication is partly guided by taking a second
look at the critique of automata theory in early formal methods, particularly
from the early process algebra literature\cite{milner, baeten},
Since much of the classic automata theory literature is not well known anymore,
the paper also provides brief historical literature survey. 
\end{abstract}

\maketitle

\section{Introduction}
The main argument of this paper is that ``classical automata theory'' provides
a sophisticated framework for modelling digital systems. This argument
would have surprised many researchers 
in the 1950s and 1960s -- they would have been surprised that anyone needed 
to make the argument. 
In a well known paper from 1959, Rabin and Scott wrote: 
 \begin{quote}\emph{
 Turing machines are widely considered to be the abstract prototype of
digital computers; workers in the field, however, have felt more and more
that the notion of a Turing machine is too general to serve as an accurate
model of actual computers. It is well known that even for simple
	 calculations it is impossible to give an a priori upper bound on the amount of tape
a Turing machine will need for any given computation. It is precisely this
feature that renders Turing's concept unrealistic.
In the last few years the idea of a finite automaton has appeared in the
literature. These are machines having only a finite number of internal
states that can be used for memory and computation. The restriction of
finiteness appears to give a better approximation to the idea of a physical
machine.} -- \cite{RabinScott}
 \end{quote}

Burks and Wang\footnote{
Burks was one of the primary engineers who built ENIAC\cite{burksprelim} and Wang
wrote the first automated theorem prover, invented Wang Tiles, 
and was the Ph.D. advisor to Stephen Cook,
 among many other accomplishments.}
 were even more ambitious in 1957.
\begin{quote}\emph{
To begin with we will consider any object or system (e.g., a physical body, a machine, an animal, or a solar system) that changes its state in time; it may or may not change its size in time, and it may or may not interact with its environment. When we describe the state of the object at any arbitrary time, we have in general to take account of: the time under consideration, the past history of the object, the laws governing the inner action of the object or system, the state of the environment (which itself is a system of objects), and the laws governing the interaction of the object and its environment. If we choose to, we may refer to all such objects and systems as automata. The main concern of this paper is with a special class of these automata: viz., digital computers and nerve nets.}
-- Burks and Wang, 1957 \cite{burks_and_wang}
\end{quote}

Somewhere in the 1970s and 1980s, the formal methods research community
reached a consensus, however, that classical automata theory was inadequate for formalizing
properties of complex software and systems. 
Robin Milner's ``Communication and Concurrency''\cite{milner} provides
the most worked out example of what went into that consensus and this paper
revisits Milner's critique of automata theory and takes a second look
at the automata theory literature to evaluate that critique\footnote{
This paper is about automata theory and has very little to say about
process algebra itself. It is not my intention to argue that process algebra
or any other formal method is better or worse than automata theory or even 
that such a comparison would make sense.}. Section \ref{sec:equivalence} examines 
Milner's critique of automata equivalences and section \ref{sec:moore}
shows how to construct automata models of concurrent systems and interaction.
Section \ref{sec:algebra} looks at functional representations of Moore
machines and briefly covers the basis of algebraic automata theory.
Section \ref{sec:background} provides 
some background on the development of the theory and a taxonomy of types of automata.

\section{Automata theory and process algebra \label{sec:equivalence}}
\subsection{Bisimulation and covering \label{sec:covering}}
In chapter 4 of ``Communication and Concurrency''\cite{milner}
Robin Milner wrote:
 \begin{quote}
\emph{We begin ... by showing the need for a notion of equality stronger than that found standard automata theory} (p 84)\end{quote}
and then on the next page:
``\emph{in standard automata theory
an {automaton} is interpreted as a language.}'' (p. 85).

Baeten\cite{baeten}  in his  2005 "History of Process Algebra" puts
it this way:
\begin{quote}\emph{
    On automata, the basic notion of equivalence is language equivalence: a behaviour
is characterized by the set of executions from the initial state to a final state.}
\end{quote}

To illustrate, Milner provides two simple 
process algebra agents which ``may be thought of as finite state automata''(p.85),
then provides two state diagrams,
\begin{center}  \includegraphics[width=0.6\textwidth]{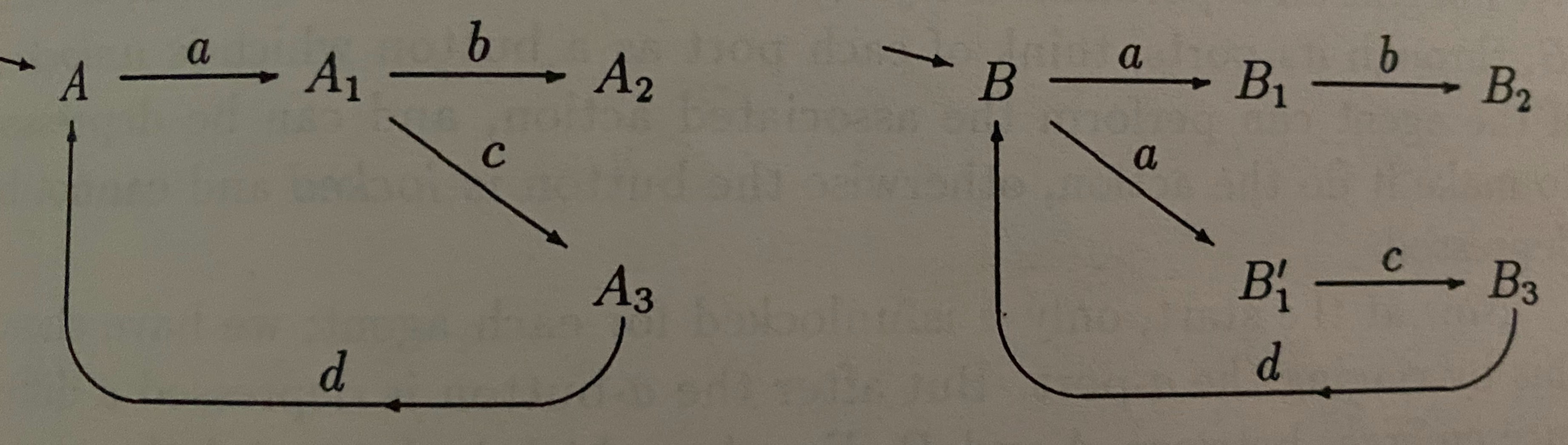} \end{center}
and explains:
\begin{quote}\emph{
if we take \(A_2\) and \(B_2\) to be the accepting states of our two automata, we can argue ... that A and B denote the same language. ...\\
But we now argue in favour of another interpretation in which A and B are different.}
\end{quote}
Milner used the same examples in a previous book \cite{milner1} where he asks "\emph{But \underline{are} they equivalent in all senses?}"  

In  automata theory A and B are definitely \underline{not} equivalent in all senses even if both accept the same language.
One is a nondeterministic finite automaton (NFA) 
and the other is deterministic (DFA), after all. They are also not equivalent
in terms of algebraic automata theory where automata can be distinguished by
behavior, structure, and even underlying semigroups\cite{ginzburg,Arbib, holcombe, HartmanisStearns}.  In a 1968 monograph, Ginzburg\cite{ginzburg}
 defines an automata relation 
called ``covering'' and a covering equivalence.
\begin{quote}\emph{
    The meaning of [B covers A] is that to every state \(s^A\) in \(S^A\) there corresponds at least one state \(s^B\in S^B\), such that when started in \(s^B\), \(B\) performs all the translations done by \(A\).
    [..]\\
    If for some A and B, B covers A and A covers B, these automata are said to be equivalent. 
    - \cite{ginzburg}, p. 97.
}
\end{quote}

In algebraic automata theory, notions like covering and machine homomorphism
were motivated by methods for factoring or partitioning automata into
simpler automata. The goal was that the product of those factors should
be equivalent, in some sense, to the original. Section \ref{sec:product}
and the sections that follow it return to this topic.

Milner's example B does not cover A (there is no B state to map \(A_1\) to) but A does cover B
so the state machines are \emph{not} equivalent in this sense.
There is no mention of any of the equivalences of 
algebraic automata theory
in ``Communication and Concurrency'' or ``Calculus of Communicating Systems''
but Milner was certainly aware of them.

\begin{figure}[bt!]
\centering
\fbox{ \begin{subfigure}{\textwidth}
\centering
\includegraphics[width=0.6\textwidth]{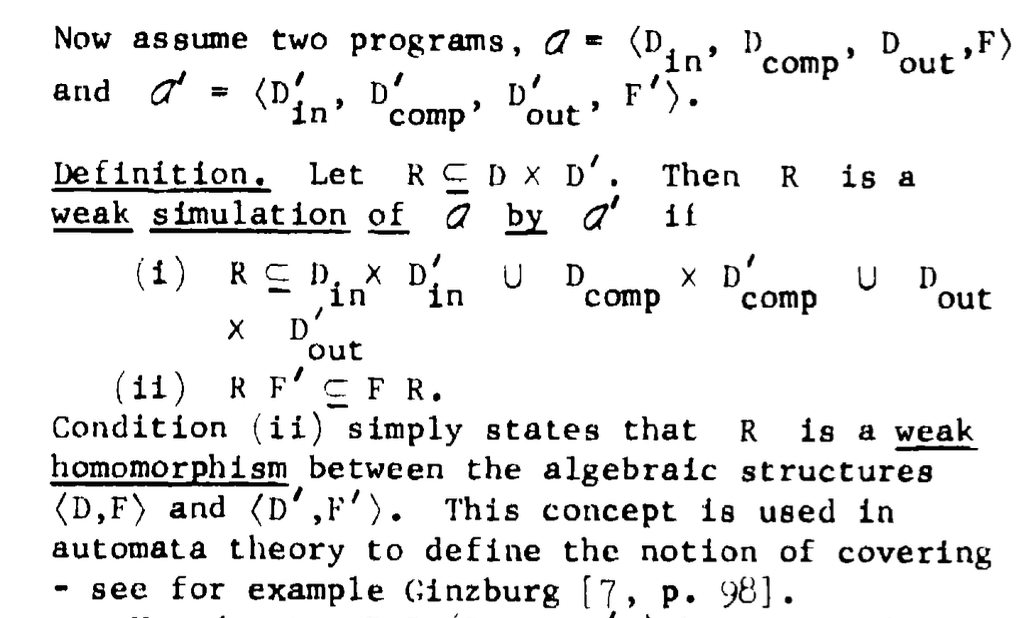}
\caption{Milner's 1971 definition of ``simulation'' \cite{milneralg}}
\label{fig:milner}
\vspace{1cm}
\end{subfigure}}
\vspace{1cm}
\fbox{\begin{subfigure}{\linewidth}
\centering
\includegraphics[width=0.8\textwidth]{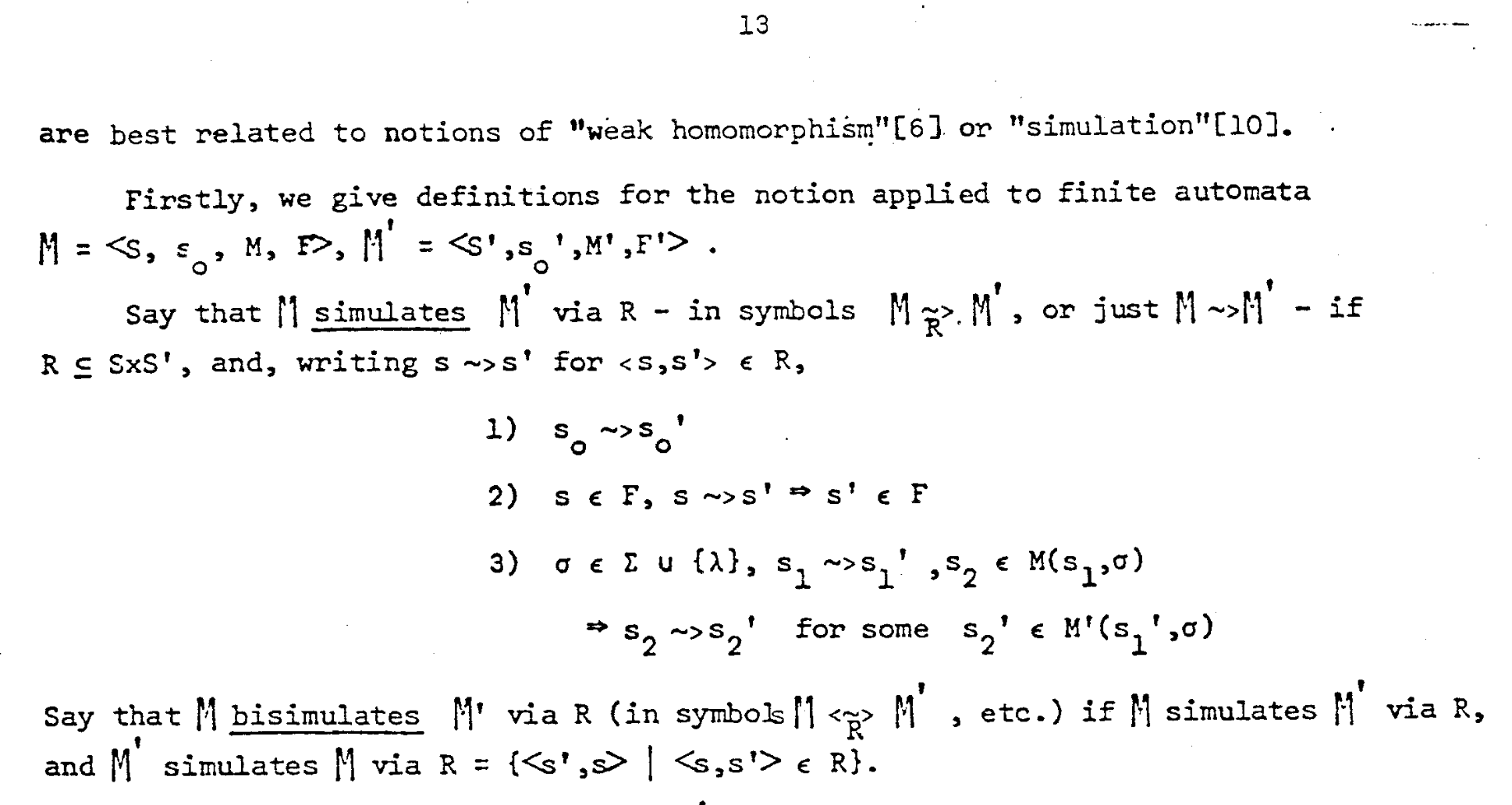}
\caption{Park's 1981 definition of bisimulation \cite{park1981} citing
Ginzburg and Milner's 1971 paper.}
\label{fig:park}
\end{subfigure}}
\caption{Origins of Bisimulation}
\end{figure}

In  a 1971 paper that preceded his work on process algebra, \cite{milneralg}
 Milner cited the same definition from Ginzburg's book:
\begin{quote}\emph{
    Condition (ii) simply states that R is a weak homomorphism between
the algebraic structures (D,F),(D',F'). This concept is used in automata
theory to define the notion of covering - see for example Ginzburg (\cite{ginzburg}, p. 98.)}
\end{quote}
See figure \ref{fig:milner} for details.
A weak homomorphism is a minor variation on covering (\cite{ginzburg}, p 99) 
with a relation in place of the map between states (the correspondence).
Park\cite{park1981},  who is generally
credited with introducing bisimulation to process algebra,
substitutes  the terms 
``simulates'' and ``bisimulates'' for Ginzburg's ``covers'' 
and ``is equivalent'' and cites both Ginzburg and Milner's 1971 paper.
\begin{quote}\emph{
The sort of rule to be discussed can be seen to develop from the known decision
procedures for problems concerning these automata- But in the form given here, they
are best related to notions of weak homomorphism\cite{ginzburg} or ``simulation'' \cite{milneralg}.
} 
[\dots]
\emph{
Say that \(M\) ``bisimulates'' \(M'\) via \(R\) [\dots] if \(M\) simulates \(M'\) via \(R\),
and \(M'\) simulates \(M\) via \(R\)}-- Park \end{quote}
See figure \ref{fig:park} for details.

Relations of the kind Ginzburg called ``cover'' are sometimes called
 ``simulations'' in the 1960s automata
theory literature (e.g. \cite{arbibchapter} p. 39).
The reader familiar with process algebra will see the similarity between
covering equivalence and process algebra
 bisimulation.  Sangiorgi\cite{sangiorgi} (p. 132)  points out that not only is bisimulation similar to automata homomorphisms
and covering
 but it is related to an even older automata theoretic concept.
\begin{quote}\emph{
	[\dots] in the 1960s weak homomorphism is well-known in
automata theory and \dots this notion is not that far from
simulation. Another emblematic example, again from automata theory, is given by
the algorithm for minimisation of deterministic automata, already known in the
1950s [Huffman \cite{huffman}; Moore 1956 \cite{moore}] (also related to this is the Myhill--Nerode theorem [Nerode 1958]).}
\par
	[\dots]\emph{
	The algorithm strongly reminds us of the Paige--Tarjan's partition refinement
algorithm [Paige and Tarjan 1987], the best known algorithm for computing bisimilarity and for minimisation modulo bisimilarity.
}\end{quote}

Sangiorgi also provides an extensive discussion of what 
differences between automata homomorphisms and process algebra bisimulation
(p. 125-129), but here our concern is automata theory and understanding 
Milner's critique. 

To illustrate, Milner describes an ``experiment''.
\begin{quote}\emph{
According to our earlier treatment of examples, A and B are agents which may interact with their environment
through the ports \(a\), \(b\), \(c\), and \(d\). We imagine an experimenter trying to interact with the agent A or with B through its ports; think of each port as a button which is \emph{unlocked} if the agent can perform the associated action and can be depressed to make it do the action, otherwise the button is \emph{locked} and cannot be depressed.}\par [...]\emph{
after the \(a\) button is depressed [ in the initial state] a difference emerges between A and B. For A -- which is deterministic --  \(b\) and \(c\) will be unlocked, while for B -- which is non-deterministic -- sometimes only \(c\) will be unlocked.} -- p. 86.
\end{quote}

The experiment works for the process algebra
 agents because their defining equations 
are transparent in terms of what events can happen in any state.
The experimenter cannot distinguish A and B \emph{as state machines} because
they do not have buttons or ports as output.
State \(B'\) does not output any information which would tell the
experimenter that the \(c\) event is the only enabled event.
In fact, the outputs of A and B are limited because
they are both a particular type of state machine called
a recognizer or an acceptor.

Rabin and Scott's 1959 paper explains the distinction
between acceptors/recognizers and automata with more general output.
\begin{quote}\emph{ ``A
neat form of the definition of automata has been used by Burks and Wang'
and by E. F. Moore, and our point of view is closer to theirs than it is to
the formalism of nerve-nets. However, we have adopted an even simpler
form of the definition by doing away with a complicated output function
and having our machines simply give "yes" or "no" answers.''}-- (page 64)\par
[...]\par
\emph{An automaton will be
considered as a black box of which questions can be asked and from which
a ```yes'' or ``no'' answer is obtained.} (page 66)
\end{quote}

The same argument, with the same examples, in Milner's earlier book
has the more illuminating title
\emph{Traditional equivalence of finite state \textbf{acceptors}} (\cite{milner1} p.9)
(with my bold added ) and the term ``standard automata theory'' only appears
in the later book. Apparently, ``standard automata theory'', for Milner,
only included these black box state machines and not machines with 
more general output, like Mealy and Moore
machines\cite{hopcroft},
 which one might otherwise consider to be part of standard automata
theory (see section \ref{sec:taxonomy} for a taxonomy of automata).

Milner has also required that
the structure of the automata not be considered. 
 \begin{quote}\emph{we only wish to distinguish between two agents P and Q
if the distinction can be detected by an external agent interacting with each of them.} -- (\cite{milner},p 84)
 \end{quote}
In his first book, he writes:
\begin{quote}\emph{
"two systems are indistinguishable if we cannot tell
 them apart without pulling them apart.''} (\cite{milner1}, p. 2).
\end{quote}
That requirement rules out the use of 
cover equivalence, machine homomorphisms, minimization,
DFA/NFA, differences and other parts of automata theory one might also
consider standard. We are only
permitted to examine the output of automata to distinguish them.

Given this combination of requirements,
Milner's opening argument could be more precisely expressed by:
\begin{quote}
\emph{We begin ... by showing the need for a notion of equality stronger than that can be supported by only examining the outputs of recognizers in 
automata theory.} \end{quote}
 But that brings up the question 
of why the agents were associated with recognizers and not automata
with more general output.
Rabin and Scott limit the automata output because they are using automata
to decide language membership, not to model devices or software. But if 
your goal is to model devices and software,
automata with more general output
are obviously more appropriate.

\section{Automata with output\label{sec:moore}}

\begin{quote}\emph{
    Basically, what is
missing [in automata theory] is the notion of interaction: during the execution from initial state to final state,
a system may interact with another system. This is needed in order to describe parallel or
distributed systems, or so-called reactive systems.}
-- Baeten\cite{baeten}
\end{quote}

Mealy and Moore machines are discussed in Hopcroft and Ullman's standard
text for computer scientists\cite{hopcroft}.
The usual definition of a Moore machine tuple is \(M=(A,Y,S,s_0,\delta,\lambda)\)
where \(A\) is a set of events, \(Y\) is a set of outputs, \(S\) is the 
state set, \(s_0\in S\) is the initial or start state,
 \(\delta:S\times A\to S\) is the transition map 
and \(\lambda:S\to Y\) is the output map. Moore machine state sets are
usually finite, in which case the automaton is called a finite state Moore
machine.

A Mealy machine differs from a Moore machine in that the output map
also depends on the input that drove the machine to its current 
state: \(\lambda:S\times A\to Y\).

Moore's 1954 paper\cite{moore} that defined what are now
called {Moore machines} describes 
``thought experiments'' that are strikingly similar to Milner's experiment.
\begin{quote}\emph{
The experimenter will choose which finite sequence of input
symbols to put into the machine, either a fixed sequence, or one in which each symbol depends on the previous output symbols. This sequence of input symbols, together with the sequence of output symbols, will be called the outcome of the experiment.}
	[...]
\emph{
	the experimenter may be thought of as a human being who is trying to learn the answer to some question about the nature of the machine or its initial state. This is not the only kind of experimenter we might imagine in application of this theory; in particular the experimenter might be another machine.} -- E.F. Moore, Gedanken-Experiments on Sequential Machines \cite{moore}
\end{quote}

Converting  Milner's \(A\) and \(B\) automata  to  Moore machines
by adding outputs to indicate enabled transitions
from the current state allows
Milner's experimenter to distinguish between the machines purely on outputs.
In a  Moore machine version of \(A\), state \(A_1\) outputs 
\(\{b,c\}\) and \(A_3\)  outputs \(\{d\}\) \emph{etc.}.  A Mealy machine
could be defined so that
every input is enabled on every step, and buttons that are ``locked''
cause a transition back to the current state and output a message ``Locked''.

Moore and Mealy machines were
introduced in the 1950s, are still widely used in digital circuits and 
software, and were central to the algebraic automata theory
literature. 
State machines with output are extensively discussed in Ginzburg's
text - the definitions of coverings and automata homomorphism
cited by Milner and Park are
for Mealy machines. On the other hand, recognizers became very important in 
computer science for definition of formal languages, parsing,
 and searching software 
(e.g. \cite{thompson1968}) and in the 1970s computer science
separated more from electrical engineering. 
By 1979 Hopcroft and Ullman's textbook  devoted just 
three out of more than four hundred pages to
``State machines with output''.
It might not have been out of the question for a
computer science researcher in the 1970s and 1980s
to think of ``standard'' automata theory 
as consisting only of recognizers. 
Edsger Dijkstra appears to have done just that in his
widely cited one line comment:
\begin{quote}\emph{
traditional automata theory tends to make us insensitive to
the role interfaces could and should play in coping with complex designs.}
\cite{EWD463}
\end{quote}

The distinction between internal state and visible state
(output) in a Moore machine
provides a rigorous definition of \emph{interface} 
which is specified by the output map and the set of input events together.
If the Moore machine is finite and in minimal form, the ratio \(\lvert S\rvert/\lvert{Y}\rvert\)
is rough indication of modularity -- how much information from the state
set is visible\cite{parnas}. Automata products even provide a model of composition and
concurrency.

\subsection{Automata products\label{sec:product}}

\begin{quote}\emph{
But meanwhile I got somehow interested, and I don't know how, in concurrency. I remember that, without linking it to verification particularly, I wondered about interacting automata. I had an idea that automata theory was inadequate, because it hadn't said what it was for two automata to interact with each other. Except for the Krohn-Rhodes Representation Theorem, which said something about feeding the output of one automata into another. But there wasn't interaction between the automata.} -- Robin Milner, interview \cite{milnerinterview}
\end{quote}

\begin{quote}\emph{Single output automata with given time delays can be combined into a new automaton} -- Von Neumann \cite{vonneumanninshannon} (p.45). 
\end{quote}

Hartmanis (1962) defined a 
\emph{concurrent product} of Moore machines \cite{hartmanis} in a well
known paper that was cited by Ginzburg. 
\begin{figure}[ht!]
\begin{center}\fbox{\includegraphics[width=0.4\textwidth]{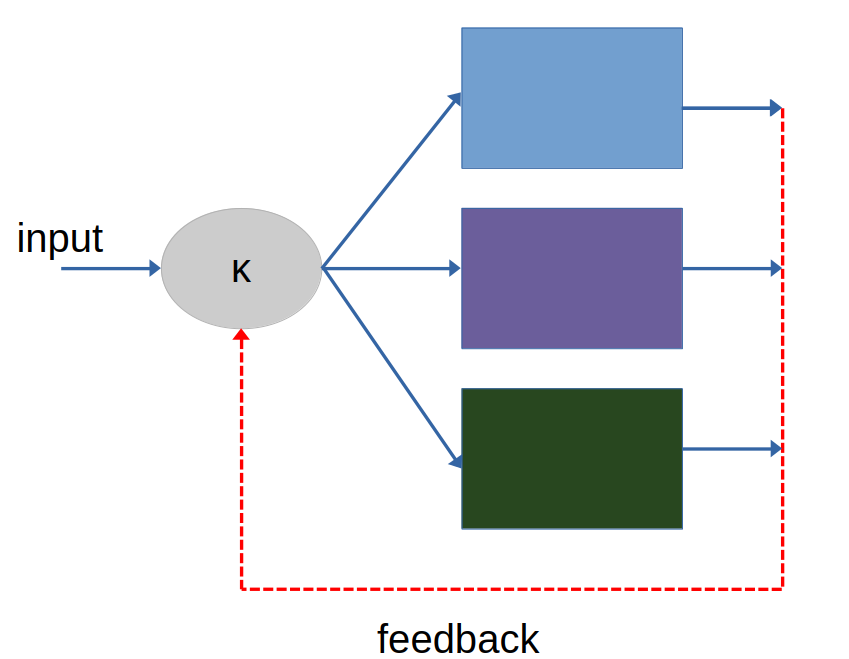}}\end{center}
\caption{The concurrent product of Moore machines\label{fig:product}}
\end{figure}

\par
\begin{quote} \emph{
DEFINITION 1. Let \(M_1, M_2, ., M_n,\) be a set of Moore type machines
	in which the outputs of any machine \(M_i, i = 1, 2,..., n\),
may be used as inputs to other machines. We shall say that this set
of interconnected machines is concurrently operating if the next
state (state at time t + 1) of each machine \(M_i\) depends on the present
state of \(M_i\), the present outputs (which depend only on the present
states) of the machines to which it is connected, and the present
external input.
The ordered n-tuple (or "configuration") of the present states of
the machines \(M_1, M_2,. , M_n\) will be referred to as the state of the
	interconnected machine.} -- Hartmanis\cite{hartmanis}, p 117.
\end{quote}

At each step, each component \(M_i\) gets input
that is a function of the external input (the input to the 
system) and the outputs of some or all of the components.
To illustrate suppose, for \(0< i\leq n\), 
\(M_i=(A_i,Y_i,S_i,s_{i,0},\delta_i,\lambda_i)\) and we want to connect
them as factors of a product state machine
\[M= (A,Y,S,(s_{1,0}\dots,s_{n,0}),\delta,\lambda).\]
Let \(A\) be the product input alphabet  (the inputs
for the system), \(Y= Y_1\dots \times Y_n\) is the output set of the product
machine and \(S= S_1\dots \times S_n\) is the state set of the product machine.
For \(s = (s_1,\dots s_n)\) let
\[\lambda(s) = (\lambda_1(s_1)\dots , \lambda_n(s_n)).\]
To complete the product, we need
a connection map:
\[\kappa:Y\times A\to A_1\dots \times A_n\] 
When the product machine is in state \(s\) an input \(a\) will advance \(M_i\)
by \(\kappa_i(\lambda(s),a)\), which is some element of \(A_i\).
That is, if \(\kappa(y,a) = (y_1,\dots ,y_n)\) then \(\kappa_i(y,a)=y_i\).
The input alphabets
of the component machines may be distinct or not, depending on what is being 
modelled. 
The product transition map \(\delta\) is given by:
\[\delta(s,a) = (\delta_1(s_1,\kappa_1(\lambda(s),a))\dots
,\delta_n(s_n,\kappa_n(\lambda(s),a)))\]

Hartmanis' version of the feedback product provides
an interesting model of composition and concurrency.
\begin{itemize}
\item Moore machine tuples are closed under this product. 
\item If all the factors are finite state the product automaton is finite state.
\item The product
models composite systems with true concurrency, not interleaving.
\item The product
does not incorporate any particular model of communication, but
shared variables, messages, shared memory, wires, or other mechanisms can be
modeled by varying the interconnection.
\item There is no need for a renaming
operation to accomplish composition
 and events with the same names in multiple event alphabets don't
have any special properties.
\item Another
property of this particular product is that it respects modularity\cite{ParnasCriteria},
each component can see only the outputs of connected components, not their
states.
\end{itemize}

Suppose we wanted to model something simple like
 synchronous message
passing with the Hartmanis concurrent product.
Then each \(A_i\) could contain a set of messages \((j,v)\) where
\(v\in V\) is a message value and \(0< j \leq n\) is the destination address
plus three additional symbols \(step,null\) and \(transmit\). The intuition
is that \(step\) is an internal calculation, \(null\) means the machine
is blocked waiting to send or receive a message or just doesn't advance
in this state for some reason with 
\(\delta_i(s_i,null)=s_i\) and \(transmit\) which means the machine has succeeded
in sending a message.  Each machine output set \(Y_i\) could contain a
set of transmit requests \((send,j,v)\) and a set of receive requests
\((receive,j)\) and a \(busy\) signal (meaning the machine just wants to 
step, if possible). Then \(\kappa\) can be partially defined by:
\[
\kappa_i(y,a) =\begin{cases} 
transmit& \text{ only if }  y_i=(send,j,v)\text{ for some }v\text{ and }y_j= (receive,i)
\\
(j,v)
&\text{ only if }y_i=(receive,j)\text{ and }y_j= (send,i,v) \text{ for some }v\\
step \text{ only if } &y_i=busy\\
null&\text{otherwise}
\end{cases}
\]

There is no
fundamental limit on how many factors can advance in parallel.
The alphabet 
\(A\) might carry information about what choices \(\kappa\) should make
and about which and how many factors can advance in one step.
Alternatively, there could be an additional factor \(M_{0}\) that would
act as a scheduler and decide which factors get inputs other than null
on each step (perhaps there are only a certain number of processor cores
available) where
\[\kappa_0(y,a) = y\]
and \(Y_0\) would provide scheduling information to \(\kappa\) as output.
The \(M_0\) machine would learn that some machine is attempting to send or
receive only one step later, but that machine would be blocked until the 
\(M_0\) machine decided to let it advance. The numbering of components
is purely arbitrary -- as long as it is consistent.

\begin{figure}[b!]
\begin{center}
\fbox{ \includegraphics[width=0.6\textwidth]{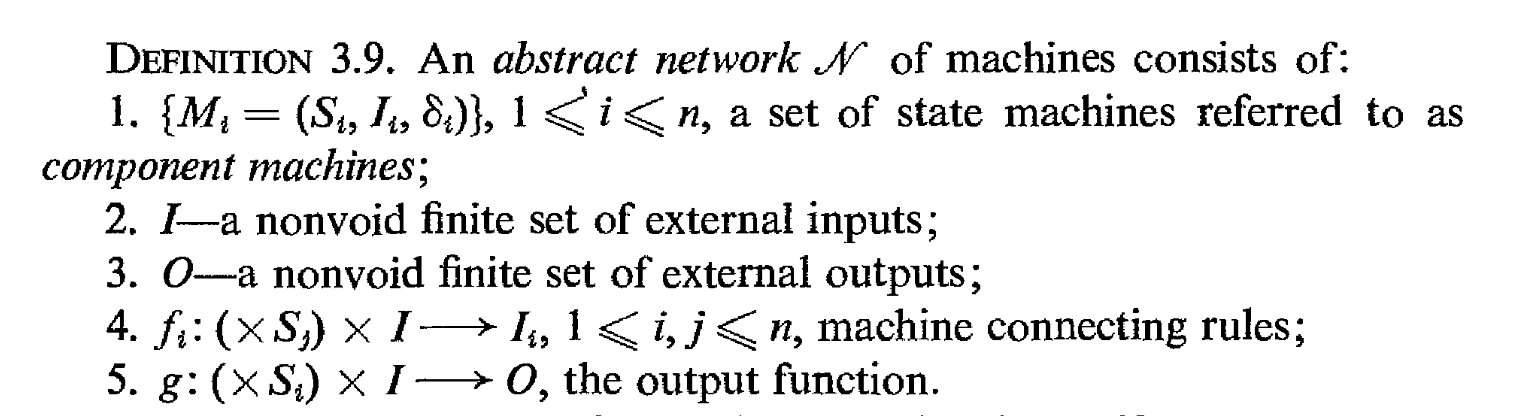}}\end{center}
\caption{Hartmanis and Stearns, 1966 \cite{HartmanisStearns}, (p 82)}
\label{fig:hartmanisstearns}
\end{figure}

Hartmanis and Stearns defined a slightly different version of the concurrent
 product 
they called a \emph{network} in their book\cite{HartmanisStearns} (
figure \ref{fig:hartmanisstearns}). Gecseg\cite{Gecseg}
has another version and 
cites other work dating back to the 1970s and earlier.
Moore in the passage quoted
above is thinking about interconnected machines where one machine would act
as the experimenter.  Interconnected
 products seem to have been considered obvious in classical 
automata theory. See
for example a passing mention in Assmus and Florentin \cite{assmus} (p. 30). 

Both Hartmanis and
Gecseg show how by limiting feedback, the interconnection can be varied.
The connector
map 
determines the connection graph.
In the cascade  product the factor automata are ordered so that
inputs generated for \(M_i\) depend only on the outputs
of \(M_1,\dots M_i\)
and ignore the outputs of the remaining factors. 

Cascade, also called loop-free, factoring of automata  was a central concern of algebraic 
automata theory in the 1960s because the successful work by Hartmanis 
and his colleagues 
using the method to simplify automata and because of the mathematical
significance 
of the relationship between loop-free products and semigroup theory
\cite{krohnrhodes} or see \cite{holcombe}. 
This relationship is briefly discussed in section \ref{sec:monoid}. 
For the purposes of modelling composite 
digital systems, however, interaction is key.

In fairness to researchers in formal methods, algebraic automata
theory appears to never have addressed the possibility
of  using these general products to  
\emph{construct} specifications of complex systems. The focus of that
work was on factoring machines that were known. 
The engineering motivation in the 1950s and 
1960s was  
simplification of what now seem like tiny circuits and devices.
By factoring the state machine, one could construct an equivalent one
as a connected sequence of simpler machines.

\section{Functional representations and  algebra}\label{sec:algebra}

\subsection{Maps\label{sec:maps}}
Each Moore machine tuple \(M\) determines a map \(M^* :A^*\to Y\) called
the \emph{characteristic map} of \(M\). Let \(\ess\) be the empty sequence
and \(w\xy a\) be the result of appending input \(a\) to finite sequence
\(w\) on the right. Then 
\[M^*(w)= \lambda(\delta^*(s_0,w))\] where
\[\delta^*(s,\ess) = s\text{ and } \delta^*(s,w\xy a) = \delta(\delta^*(s,w),a)\]
This map captures the behavior of \(M\).

Arbib \cite{Arbib} (1968) writes
\begin{quote}\emph{ ``we may say automata theory is the study of partial functions \(F:A^*\to Y^*\)''} -- (p. 6).
\end{quote}
Similarly, Pin\cite{pin} (p. 100) defines a sequential function \(\sigma:A^*\to Y^*\) as 
``a function whose behavior is defined by a machine called a `sequential transducer' ''. 
The set names on both of these are changed for consistency here.
This map is easily obtained from the characteristic map defined above by:
\[M^{**}(\ess)= \ess\text{ and } M^{**}(w\xy a) = M^{**}(w)\xy M^{*}(w)\]

Variations
of these maps
can be found all over the automata theory literature (e.g. \cite{assmus,ginzburg,holcombe,pin}).
In \cite{assmus} (p. 17) they are called input-output functions. These
maps are primitive recursive in the sequence \cite{petercomputer} but either
automata theory researchers did not recognize these maps
as primitive recursive or they
didn't consider it important enough to mention. Peter's 
book which covers primitive recursion on finite sequences
does not mention state machines.  But primitive recursive maps on finite
sequences provide a representation of Moore machines  and products of Moore machines that can
be used as an alternative to state diagrams and 
Moore machine tuples. This approach is the subject of \cite{yodaikenlarge}.
and has advantages for scaling Moore machines to
large state sets and complex systems to address the objections to state
diagrams given by Harel\cite{statechart} (p 233-234).

\subsection{Algebra\label{sec:monoid}}

The relationship between automata, automata products, and factoring
semigroups and monoids is extensively discussed in algebraic automata
theory. See \cite{holcombe} for an introduction.

There are several similar
 ways to construct a monoid from a state machine. One way is to use
the input/output maps of the previous section and define a
congruence.

\[\begin{array}{l}
\text{For }w,u\in A^* \text{ define } (w = u \bmod M) \text{ if and only if }\\
\quad (\forall z_1,z_2\in A^*)(M^*(z_1\concat w\concat z_2) = 
M^*(z_1\concat u\concat z_2))\end{array} \] 

Then the elements of the monoid are the equivalence classes 
\([w]_M = \{u\in A^*: u \sim w \bmod M\}\). The monoid
operation is \([w]_M\circ [u]_M = [w\concat u]_M\) and the identity element
is \([\ess]_M\).

If a monoid has a cancelation property, it is a group. An automaton with 
a monoid that is a group is called a group machine. 
Krohn-Rhodes theorem \cite{krohnrhodes} surprisingly ties factoring 
of finite automata using loop-free (cascade)
 products to the Jordan-Holder theorem of group theory\cite{maclane} (p. 430-432). As summarized
by Zeiger\cite{zeigerchapter}, one consequence of Krohn-Rhodes is:
\begin{quote}\emph{ Each finite state automaton can be built as a cascade of two
state automata and simple-group automata}
(p. 77)\end{quote} 

The general product of section \ref{sec:product} does not preserve the 
group structure of automata. In fact, if \(M\) has \(n\) states, it can
be factored to a product of \(\lceil \log_2 n \rceil\) two state automata.
In the general case, each transition of the product
transmits the one bit state of every factor to every other factor. 
These factorizations can be limited either by limiting the
connectivity of the connection graph or by limiting the quantity of information carried by the connection map (the bandwidth) but there doesn't seem to be
much work on either beyond what was known in the 1960s. See Gecseg \cite{Gecseg}
 for a survey.

\section{Background on state machines}\label{sec:background}
\subsection{Scope of automata theory}

Von Neumann \cite{vonneumanninshannon} wrote:
\begin{quote}\emph{
It has been pointed out by A.M.Turing in 1937 and by W.S. McCulloch ad W. Pitts in 1943 that effectively constructive logics, that is, intuitionistic logics, can be best studied in terms of automata.} (p. 44).
\end{quote}

Both Rabin and Scott and Von Neumann 
\cite{vonneumanninshannon} trace automata theory back to nerve nets\cite{kleene} but
a case could be made that the origin was mostly in switching theory.

Much of the work in automata theory was in the area
of digital circuit design  originally for design of telephone switches,
 and it was common to see articles on relays in the same conferences and 
journals that presented articles on topics like formal languages, logic,
and semigroups. Huffman's paper \cite{huffman} seems to have been the first
to develop the notion of a sequential 
circuit (a state machine with output). 
\begin{quote}\emph{ 
We may generalize from these two simple examples: In a circuit
having no secondary relays there can be no ``memory''; the states of
operation of the primary relays uniquely determine the output transmissions. Such a circuit is called a combinational circuit. In a circuit
having secondary relays, the possibility of a ``memory'' exists since the
states of operation may not uniquely determine the output transmissions. A circuit having secondary relays will be called a sequential
circuit.} - Huffman.
\end{quote} 
See also \cite{toma} for an account of the origins of switching theory.

An amazing
volume edited by Claude Shannon and John McCarthy\cite{shannon} gives some sense of the field in the mid 1950s.

\subsection{Types of automata\label{sec:taxonomy} }
Mostly following Ginzburg's taxonomy the types of state machines are as follows.
\begin{itemize}
\item A state machine or subautomaton 
	consists of a tuple \((A,S,s_0)\) where 
\(A\) is the event (input)
alphabet, \(S\) is a set of states with \(s_0\in S\)
the start state,\(\delta:S\times A \to S\) is 
the transition function.
\item A recognizer\cite{RabinScott}
adds a set \(F\subset S\), the set of accept states.
This machine accepts or
recognizes a finite sequence of inputs  if following it 
via \(\delta\) from the start state through the 
end of the sequence terminates at a state in \(F\).
\item A Moore machine tuple\cite{moore} \((A,S,s_0,X,\lambda)\) replaces
the set of accept states with a set \(X\) of outputs
and a map \(\lambda:S\to X\). Moore machines can act as acceptors when
the output map is suitable.
\item A Mealy machine\cite{mealy} \((A,S,s_0,X,\gamma)\)  replaces
	the Moore machine output map \(\lambda\)
		with some \(\gamma:S\times A\to X\).
\item \emph{Sequential machine} is a generic term, usually used
more in circuit design which usually has some output see
		Huffman \cite{huffman} and
 Burks and Wang\cite{burks_and_wang}
\end{itemize}

		Turing machines are also often considered
		automata  as are linear bounded automata
		\cite{hopcroft} and Buchi automata
		among others. Those are
		not discussed here.

\bibliographystyle{alphaurl}
\bibliography{milner.bib}
\end{document}